\documentclass[preprint,prd,
amsmath,amssymb,amsthm,nofootinbib]{revtex4}
\usepackage[mathscr]{euscript}
\usepackage{threeparttable}
\usepackage{subfigure}
\usepackage{bm}
\usepackage{textcomp}
\usepackage{graphicx}
\usepackage{subfigure}
\usepackage{overpic}
\usepackage{multirow}
\usepackage{floatrow}
\usepackage{float} 
\usepackage{amsmath,amssymb,amsthm}
\usepackage{cases}
\usepackage[colorlinks=true,linkcolor=red]{hyperref}
\newfloatcommand{capbtabbox}{table}[][\FBwidth]

\begin{document}
	
	\title{ Alleviating the Hubble-constant tension and the growth tension via a transition of absolute magnitude favored by the Pantheon+ sample}
	
	\author{Yang Liu$^{1}$\footnote{yangl@hunnu.edu.cn}, Hongwei Yu$^{1,2}$\footnote{Corresponding author: hwyu@hunnu.edu.cn} and Puxun Wu$^{1,2}$\footnote{Corresponding author: pxwu@hunnu.edu.cn}}
	\affiliation{
		$^{1}$Department of Physics and Synergetic Innovation Center for Quantum Effects and Applications, Hunan Normal University, Changsha, Hunan 410081, China \\
		$^{2}$Institute of Interdisciplinary Studies, Hunan Normal University, Changsha, Hunan 410081, China
		}

\begin{abstract}

We establish a cosmological-model-independent method to extract the apparent magnitude and its derivative at different redshifts from  the Pantheon+ type Ia supernova sample, and find that the obtained values deviate clearly from  the prediction of the $\Lambda$CDM model at the lowest redshift.  This deviation can be explained  as  a result of a  transition of the  absolute magnitude $M$ in the low redshift region. The observations seem to favor this transition since the minimum values of $\chi^2$ for two ansatzes of a varying $M$   are less than that of a constant $M$.  The Hubble constant tension is alleviated from larger than $5\sigma$ to be about $1$ to $2$$\sigma$ for a varying $M$, and the growth tension can be resolved after attributing  the variation of $M$ to a modification of the effective Newton's constant.

\end{abstract}
\maketitle
	
\section{Introduction}

The cosmological constant $\Lambda$ plus cold dark matter ($\Lambda$CDM)  is the simplest and  most popular   cosmological model.  It is well consistent  with many observations on one hand,  but on the other hand,  it  still  suffers  the serious Hubble constant ($H_0$) tension~\cite{Perivolaropoulos2022,Riess2020,Tully2023}, which  refers to a more than 5$\sigma$ disagreement between the measurements of $H_0$ given respectively  by the SH0ES Collaboration~\cite{Riess2022} and the Planck satellite~\cite{Planck2020}. Within the framework of  the $\Lambda$CDM model, the cosmic microwave background (CMB) radiation data from the Planck satellite  infer $H_0=67.4\pm 0.5~\mathrm{km~s^{-1}~Mpc^{-1}}$~\cite{Planck2020}, which deviates significantly from $H_0=73.04\pm 1.04~\mathrm{ km~s^{-1}~Mpc^{-1}}$ constrained  cosmological-model-independently  by the data from  the nearby type Ia supernovae (SNe~Ia)~\cite{Riess2022}.  These SNe Ia are calibrated by  using  the Cepheids  according to the  idea of distance-ladder,  and   the absolute magnitude $M$ of SNe~Ia  is determined to be $M=-19.253\pm0.027~ \mathrm{mag}$. 
To figure out whether the $H_0$ tension  originates from the calibration  of SNe Ia, the Mira variables have been used to calibrate the SNe Ia,  resulting in $M=-19.27\pm0.13~ \mathrm{mag}$ which yields  $H_0=72.7\pm4.6~ \mathrm{km~s^{-1} Mpc^{-1}}$~\cite{Huang2020}. This result is consistent with that  from the Cepheid-calibrated SNe Ia; while it is larger than   $H_0=69.8\pm1.7~\mathrm{km~s^{-1} Mpc^{-1}}$ and $70.50\pm4.13~\mathrm{km~s^{-1} Mpc^{-1}}$ obtained respectively from the SNe Ia calibrated with  the tip of the red giant branch~\cite{Freedman2021} and the surface brightness fluctuations~\cite{Khetan2021}.  However, if  the idea of inverse distance-ladder and  the high redshift data, such as the baryon acoustic oscillation, are utilized to calibrate the SNe Ia, a value of $M$ smaller than that from the Cepheids and a value of $H_0$ consistent with that  from the Planck CMB data are achieved~\cite{Macaulay2019,Abbott2018}. Apparently,   a smaller $M$ seems to give a smaller $H_0$. Thus,   the $H_0$ tension can also be regarded as the $M$ tension~\cite{Camarena2020}.   

The $H_0$   tension may be caused by either systematic errors or local bias. Unfortunately, no systematics, which  could explain this tension, have been found so far~\citep{Efstathiou2014,Feeney2018,Riess2016,Cardona2017,Zhang2017,Follin2018,Riess2018a,Riess2018b,Huillier2019,Wagner2019,Wang2024,Benisty2023,Castello2022}, and a  local void  cannot save the tension either~\cite{Kenworthy2019,Lukovic2020,Cai2021,Hu2024}. Therefore, the $H_0$ tension may be the smoking gun of new physics beyond the $\Lambda$CDM model either in the early or late universe~\cite{Verde2019}. 
A simple extension of the $\Lambda$CDM model in the late universe  is to replace the cosmological constant with a dynamical dark energy, such as that described by the the Chevalier–Polarski–Linder (CPL) parameterization. However, these extensions cannot fully solve the tension~\cite{Guo2019,Okamatsu2021,Alestas2022,Hu2023} since they  only enlarge the uncertainties of  the constraints on the cosmological parameters.  Noteworthily,  reducing the cosmic sound horizon, which can be realized by modifying the recombination history or introducing an early dark energy~\cite{Jedamzik2020,Hart2020,Sekiguchi2021,Poulin2019}, seems to be capable of resolving the $H_0$ tension, but  
it may regrettably worsen  the so-called growth  tension at the same time~\cite{Jedamzik2021,Hill2020}. This tension refers to  the about $3\sigma$ disagreement between  the values of the matter density parameter $\Omega_\mathrm{m0}$ and the parameter $\sigma_8$ constrained, respectively,  from   the Planck 2018 CMB data~\cite{Planck2020} in the $\Lambda$CDM background geometry  and the dynamical probes of the cosmological perturbations including cluster counts~\cite{Rozo2010,Rapetti2009,Bocquet2015,Ruiz2015}, weak lensing~\cite{Schmidt2008,Hildebrandt2017,Heymans2012,Joudaki2018,Troxel2018,Kohlinger2017,Abbott2018a,Abbott2019} and redshift-space distortions~\cite{Samushia2013,Macaulay2013,Johnson2016,Nesseris2017,Kazantzidis2018}. Here $\sigma_8$ is defined as the matter density rms fluctuations  in spheres of radius $8h^{-1}~\mathrm{Mpc}$ at $z=0$ with $h \equiv \frac{H_0}{100~\mathrm{ km~s^{-1}Mpc^{-1}}}$. Therefore,  the Hubble constant  tension remains to be an open issue in modern cosmology.

A possible way to find out the origin of the $H_0$ tension is to probe directly the cosmic background dynamics from the observational data.  In this paper, we propose a model-independent method to extract the apparent magnitude $m$ and its derivative $m'=\frac{dm}{dz}$  at  different redshift points from the latest Pantheon+ SNe Ia sample~\cite{Scolnic2022}.    We find that except for the results at the lowest redshift point, the obtained values of $m$ and $m'$ are very well compatible   with the prediction from the $\Lambda$CDM model.  Thus, it is reasonable to  assume that the $\Lambda$CDM model can describe   correctly the cosmic evolution, and  the deviation of $m$ and $m'$ from the prediction of the $\Lambda$CDM model at the low redshift region originates from a transition of the absolute magnitude $M$ of SNe Ia. We demonstrate that such a transition of $M$ will alleviate the $H_0$ tension. If  the transition of $M$ is further assumed to arise from the variation of the effective Newton's constant $G_\mathrm{eff}$, the growth tension can be resolved too.

\section{Values of apparent magnitude and its derivative}\label{sec:2}

One observable of SNe Ia is the apparent magnitude $m(z)$. Its theoretical value relates to the cosmological model through 
\begin{eqnarray}\label{eq:m}	m_\mathrm{th}(z)&=&25+5\log_{10}\left(\frac{D_L(z)}{\mathrm{Mpc}}\right)+5\log_{10}\left(\frac{c}{H_0}\right)+M. 
\end{eqnarray}
Here $c$ is the speed of light,   and $D_L(z)$ is the dimensionless luminosity distance, which is defined as $D_L(z)\equiv (1+z)\int_0^z \frac{dz'}{E(z')}$   in a spatially flat universe, where $E(z)$ is the dimensionless Hubble parameter and $E(z)\equiv\sqrt{\Omega_\mathrm{m0}(1+z)^3+(1-\Omega_\mathrm{m0})}$  for the $\Lambda$CDM model. Comparing the observed $m(z)$ with its corresponding theoretical value can give  constraints on the cosmological models with the SNe Ia data, {\it e.g.}, $\Omega_\mathrm{m0}=0.333 \pm 0.018$ in the $\Lambda$CDM model with the Pantheon+ SNe Ia sample, which comprises 1701 light curves with 1550 distinct SNe~Ia, and spans to redshift  $z\simeq2.26$~\citep{Scolnic2022}. If  a prior on $M$ is further given, a constraint on $H_0$ will be achieved by using SNe Ia.  
 With $M=-19.253\pm0.027~\mathrm{mag}$  from the Cepheids, the Pantheon+ SNe~Ia sample gives $H_0=73.22\pm0.95~ \mathrm{km~s^{-1}Mpc^{-1}}$ in the $\Lambda$CDM model.

To model-independently probe the local background dynamics of our universe with the Pantheon+ sample, we now establish  a local expansion method, which is to expand the apparent magnitude $m(z)$ at a given redshift. 
 We do the Taylor expansion of $m(z)$ in the $\ln z$ space instead of the $z$ space, to the first order:
\begin{eqnarray}\label{mz}
m(z)=m_i+z_i m'_i (\ln z-\ln z_i), \quad \mathrm{if} ~~ z_{\mathrm{min},i}<z\leq z_{\mathrm{max},i},
\end{eqnarray}
where $m_i\equiv m(z_i)$, $m'_i\equiv\frac{dm}{dz}\arrowvert_{z=z_i}$, and $z_i$ is the redshift point where the expansion is performed, which is determined by using $\ln z_i=(\ln z_{\mathrm{min},i}+\ln z_{\mathrm{max},i})/2$ in our analysis.

We consider the  Pantheon+ SNe Ia  sample,  and use the Hubble diagram redshift $z_\mathrm{HD}$, which is derived from the CMB frame redshift ($z_\mathrm{CMB}$) with corrections from  the peculiar velocity,  as the redshift $z$ of the Pantheon+ sample. We exclude those data points whose redshifts are less than 0.01 since  the nearby  sample may be impacted by their peculiar velocities~\citep{Brout2022}. Furthermore, we also ignore the data with the redshift  $z>0.8$ since only 30 data points lie in the redshift region $z\in (0.8, 2.26]$. Thus, the remaining 1560 data points are used in our analysis. We divide these data into five bins with the same number of data points in each bin. 
As there are two free parameters ($m_i$ and $m'_i$) in each bin, we have totally ten free parameters. These parameters 
are constrained by  minimizing the following $\chi^2$  
\begin{eqnarray}\label{eq:chi}
	\chi^2=[\bm{\hat{m}}_\mathrm{obs}-m(z)]^\dagger C^{-1}_\mathrm{SN} [\bm{\hat{m}}_\mathrm{obs}-m(z)]
\end{eqnarray} 
from the Pantheon+ SNe Ia data. Here  $C_\mathrm{SN}$ is the covariance matrix of $1560\times1560$, which is a submatrix of the full SNe~Ia sample, and $\bm{\hat{m}}_\mathrm{obs}$ is the 1D array consisting of the SNe Ia apparent magnitudes. 
Ten free parameters, i.e., $m_i$ and $m'_i$ with $i$ varying from 1 to 5, are simultaneously fitted  by using the Markov Chain Monte Carlo (MCMC) method. Before using the real data to constrain these free parameters, we need to check the reliability of our method.  To do so,  we first mock 1560 SNe Ia data points in the redshift region of $0.01\leq z\leq0.8$ with the  value of $\langle m_\mathrm{th}\rangle$ from the fiducial model: the $\Lambda$CDM model  ($\Omega_\mathrm{m0}=0.333$, $H_0=73.22~ \mathrm{km~s^{-1}Mpc^{-1}}$, and $M=-19.253~\mathrm{mag}$), and   the same redshift distribution as that of the  Pantheon+ sample.   The mock data are divided into four, five, and six bins with the same number of points in each bin, respectively. Then, the best fitting values of $m_i$ and $m'_i$ in each bin from the mock data can be estimated by using the minimum $\chi^2$ method. We repeat our analysis 1000 times, and find that the mean values of $m_i$ and $m_i'$ are well consistent with those derived from the fiducial model for the cases of five and six bins. Thus,  the simulation analysis  shows that the results from real data will be reliable if the bin number is larger than four. The detailed discussions can be found in  Appendix \ref{sec:app}.

Table~\ref{tab:1} lists the constraints on  $m_i$ and $m'_i$ in each bin, and on $\Delta m_i\equiv m_i-m_{i,\mathrm{th}}$ and $\Delta m'_i\equiv m'_i-m'_{i,\mathrm{th}}$, which  represent the differences between the values from the Pantheon+ sample and the prediction of the fiducial model.  It is easy to see that in the last four bins, the constraints on $m_i$ and $m'_i$ are  very well consistent with those of the fiducial model. However, in the first bin ($z_1=0.017$), the value of  $\Delta m_1$ is compatible with zero at 2$\sigma$ confidence level (CL), whereas $\Delta m'_1$ deviates from zero  at about 2.7$\sigma$. 

For a more comprehensive comparison between the observed and simulated datasets, we extended our analysis to include cases with four and six bins.  The results obtained from the Pantheon+SNe Ia data are presented in Table~\ref{tab:3} and Table~\ref{tab:4} in Appendix~\ref{sec:app2} respectively. In both cases, the values of $\Delta m'_1$ from the observed data  are inconsistent with zero at more than $2\sigma$ CL. This  result is different from what is obtained from the mock data, but it is similar to the  five bin result. And all other results are compatible with the prediction of the fiducial model at $2\sigma$ CL. 

We also study the possible volume effect in the redshift region $0.01<z\leq0.027$, and find that it can not fully account for the deviation of  $\Delta m_1$ and $\Delta m'_1$.
The volume effect here refers to the bias on the low redshift Hubble diagram of SNe Ia caused by the   peculiar velocities of high redshift SNe Ia host galaxies.  This bias arises because the number density of galaxies per unit distance generally increases as the square of distance. Consequently, the number density of SNe Ia per unit distance at higher redshifts is larger than that at lower redshifts. Therefore, more SNe Ia located at higher redshifts and within a greater volume will be scattered down to lower redshifts under the influence of their host galaxies' peculiar velocities, compared to the reverse scenario~\cite{Brout2022,Kenworthy2022,Perivolaropoulos2023}. 
To clearly demonstrate the impact of peculiar velocities on our analysis, we consider the Pantheon+ Type Ia supernovae (SNe Ia) sample, using the CMB frame redshift ($z_\mathrm{CMB}$) instead of the Hubble diagram redshift ($z_\mathrm{HD}$). This sample comprises 1558 data points within the redshift range of $0.01 < z < 0.8$. The findings are detailed in the lower part of Table~\ref{tab:1}. When comparing results derived using $z_\mathrm{CMB}$ with those using $z_\mathrm{HD}$, we observe that the constraints on parameters $m_i$ and $m'_i$ for $i \geq 3$ remain largely unaffected by the choice of redshift, consistently aligning with the predictions of the fiducial model. However, the parameters $\Delta m_2$ and $\Delta m'_2$, which align with zero within $1\sigma$ CL when using $z_\mathrm{HD}$, deviate from zero beyond $1\sigma$ CL with $z_\mathrm{CMB}$, with the deviation of $\Delta m'_2$ reaching $2.24\sigma$. Additionally, while the deviation of $\Delta m_1$ from zero becomes more significant, increasing from just over $1\sigma$ to more than $2\sigma$, the value of $\Delta m'_1$ remains similar to that obtained using $z_\mathrm{HD}$. 
This analysis highlights that the deviations of $\Delta m_1$ and $\Delta m'_1$ from zero are robust to the choice of redshift.
Therefore, our results show that in the low redshift region the Pantheon+ SNe~Ia data supports the deviation of  cosmic evolution from the prediction of the $\Lambda$CDM model.

\begin{table}
	\caption{\label{tab:1}
		\footnotesize
		 Expanding Redshift Point $z_i$, Number of SNe~Ia, and Constraints on $m_i$ and $m'_i$.
	}
	\centering
	\scriptsize
	\begin{threeparttable}
		\begin{tabular}{c|ccccc}
			\hline
			\hline
			\multicolumn{6}{c}{Pantheon+ Sample with $z_\mathrm{HD}$}\\
			\hline
			& bin 1 & bin 2 & bin 3 & bin 4 & bin 5 \\
			\hline
			$z_i$ & 0.017 & 0.049 & 0.144 & 0.296 & 0.544 \\
			~Redshift range~ & ~$0.010<z\leq0.027$ ~ & ~ $0.027<z\leq0.087$ ~ & ~ $0.087<z\leq0.237$ ~ & ~ $0.237<z\leq0.370$ ~ & ~ $0.370<z\leq0.799$  \\
			$N_\mathrm{SN}$ & 312 & 312 & 312 & 312 & 312 \\
			\hline
			$m_i$ & $14.955\pm0.015$ & $17.338\pm0.009$ & $19.800\pm0.010$ & $21.563\pm0.008$ & $23.088\pm0.011$ \\
			$m'_i$ & $125.735\pm2.571$ & $45.468\pm0.511$ & $16.530\pm0.223$ & $8.543\pm0.202$ & $4.704\pm0.076$ \\
			\hline
			$\Delta m_i$  & $ {0.024\pm0.016}$ & $0.006\pm0.011$ & $-0.003\pm0.013$ & $0.006\pm0.014$ & $-0.017\pm0.020$\\
			$\Delta m'_i$  & $ {-6.894\pm2.571}$ & $-0.463\pm0.512$ & $0.084\pm0.224$ & $0.146\pm0.203$ & $-0.072\pm0.079$\\
			\hline
			\hline
			\multicolumn{6}{c}{Pantheon+ Sample with $z_\mathrm{CMB}$}\\
			\hline
			$z_i$ & 0.016 & 0.049 & 0.146 & 0.296 & 0.544 \\
			~Redshift range~ & ~$0.010<z\leq0.027$ ~ & ~ $0.027<z\leq0.090$ ~ & ~ $0.090<z\leq0.237$ ~ & ~ $0.237<z\leq0.371$ ~ & ~ $0.371<z\leq0.799$  \\
			$N_\mathrm{SN}$ & 312 & 312 & 312 & 311 & 311 \\
			\hline
			$m_i$ & $14.957\pm0.015$ & $17.378\pm0.009$ & $19.846\pm0.010$ & $21.564\pm0.008$ & $23.093\pm0.011$ \\
			$m'_i$ & $126.639\pm2.606$ & $44.275\pm0.497$ & $16.228\pm0.212$ & $8.484\pm0.197$ & $4.686\pm0.075$ \\
			\hline
			$\Delta m_i$  & $ {0.037\pm0.016}$ & $0.019\pm0.012$ & $0.000\pm0.013$ & $0.006\pm0.014$ & $-0.013\pm0.019$\\
			$\Delta m'_i$  & $ {-6.689\pm2.606}$ & $-1.114\pm0.498$ & $0.048\pm0.214$ & $0.083\pm0.198$ & $-0.091\pm0.078$\\
			\hline
		\end{tabular}
		\begin{tablenotes}
			\item[a] The mean values with 1$\sigma$ uncertainty are shown.
			\item[b]  $\Delta m_i$($\Delta m'_i$) denote the differences between the constraints on $m_i$($m'_i$) and the fiducial model: $\Lambda$CDM model with $\Omega_\mathrm{m0}=0.333\pm 0.018$ and $\mathcal{M}=25+5\log_{10}\left(\frac{c}{H_0}\right)+M=23.808\pm0.007$. 
		\end{tablenotes}
	\end{threeparttable}
\end{table}

\section{a variation of absolute magnitude}\label{sec:3}
We have found that the  $\Lambda$CDM model is inconsistent with the SNe Ia observations only in the low redshift region. Thus, it seems to be a reasonable assumption that  the $\Lambda$CDM model provides a correct description of the cosmic evolution. Then,  Eq.~(\ref{eq:m}) indicates that the discrepancy  between the values of the apparent magnitude  from observations and the prediction of the $\Lambda$CDM model  may originate from a variation of the absolute magnitude $M$. 

We now first consider a simple ansatz that $M$  varies suddenly by an amount of  constant $A$ at redshift $z_t$, {\it i.e.}, 
\begin{eqnarray}\label{eq:mod_M1}
	 M(z)=\left\{\begin{array}{ll}
		M_0 & \mbox{ if } z< z_t \\
		M_0+A & \mbox{ if } z\geq z_t \ ,
	\end{array} \right.
\end{eqnarray}
where $M_0$ is the absolute magnitude of SNe Ia calibrated from the distance-ladder, {\it i.e.}, the Cepheids, and thus can be set  to be $M_0=-19.253 \pm 0.027 $~mag. Substituting Eq.~(\ref{eq:mod_M1}) into Eq.~(\ref{eq:chi}), we find that all $m_i$ and $m_i'$ will be consistent with the prediction of the $\Lambda$CDM model.

\begin{figure}
	\centering
	\includegraphics[width=0.6\textwidth]{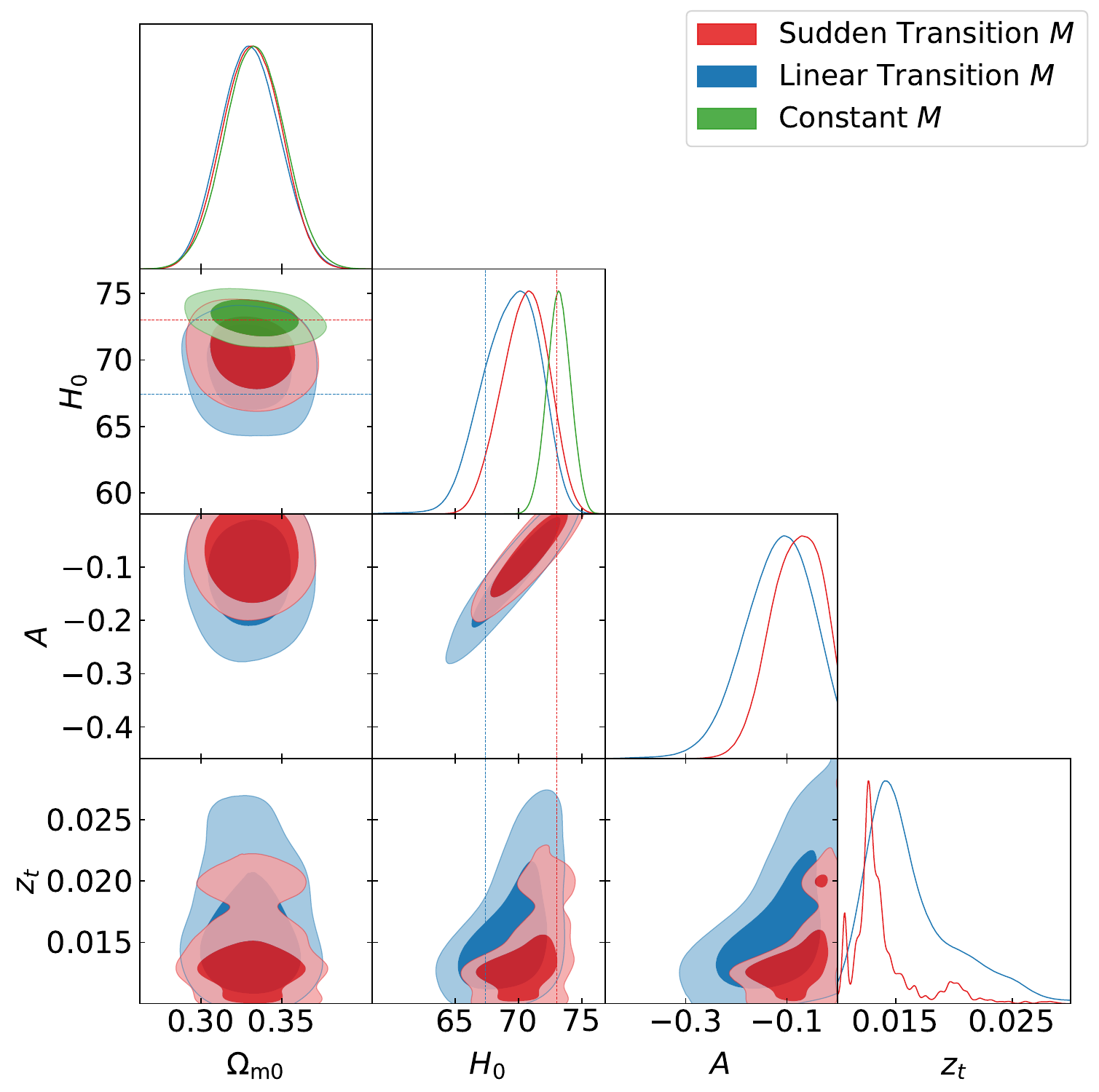}
	\caption{Constraints on cosmological parameters and parameters describing the variation of $M$ from the Pantheon+ SNe~Ia data.  
	The blue and red dotted lines show the constraints on $H_0$ from the Planck CMB and SH0ES, respectively.
		\label{Fig:1}}
\end{figure}

Next,  we study  the constraints  on $H_0$, $\Omega_\mathrm{m0}$, $A$ and $z_t$ with the Pantheon+ SNe Ia data  using the MCMC method. Although  parameter $z_t$ only exists under the condition of the piecewise function and does not appear explicitly in Eq.~(\ref{eq:mod_M1}), it is treated as a variable in our analysis since the triad $\{H_0, \Omega_{\rm m0}, A\}$ are constrained for different $z_t$ choices.  
Thus, the values of $H_0$, $\Omega_\mathrm{m0}$, $A$ and $z_t$ are sampled simultaneously.
The results are shown in Fig.~\ref{Fig:1} and Table \ref{tab:2}. The best fitting  values  are $\Omega_\mathrm{m0}=0.331$, $H_0=69.08~\mathrm{ km~s^{-1}~Mpc^{-1}}$, $A=-0.129$, and $z_t=0.0126$, with $\chi^2_\mathrm{min}=1393.3$.
Their mean values with $1\sigma$ uncertainty are   $\Omega_\mathrm{m0}=0.332\pm0.018$, $H_0=70.5^{+2.0}_{-1.7}~\mathrm{ km~s^{-1}~Mpc^{-1}}$, $A=-0.084^{+0.061}_{-0.038}$, and $z_t=0.0139^{+0.0003}_{-0.0035}$, respectively.
If  $M=M_0$, we find $\Omega_\mathrm{m0}=0.333\pm 0.018$ and $H_0=73.22\pm0.95$ with 
$\chi^2_\mathrm{min}=1402.1$.  Apparently, when $M$ varies as shown in Eq.~(\ref{eq:mod_M1}), the minimum of $\chi^2$ decreases by an amount of about $8.8$. The variation however has negligible impacts on the constraint on $\Omega_\mathrm{m0}$. The mean value of $z_t$ shows that the transition of $M$ occurs in the redshift region between $0.010$ and $0.027$, which is consistent with the result obtained in the previous section. The SNe Ia data favor a value of $M$ smaller  than $M_0$ at the redshift region $z\geq z_t$ since $A$ is negative and deviates from zero at more than $1\sigma$ CL, which results in the value of $H_0$ smaller than $H_0=73.04\pm1.04~\mathrm{ km~s^{-1}~Mpc^{-1}}$ from the SH0ES collaboration. Although the value of $H_0$ from the SNe Ia with a sudden variation of $M$ still deviates slightly  from that from the CMB data, this deviation reduces to be about 2$\sigma$ CL. Thus, a sudden decrease of $M$ in the low redshift region will alleviate the $H_0$ tension.   We must point out that a transition of the SNe~Ia absolute magnitude from a large value to a small one at low redshift ($z \simeq 0.01$) was  first proposed in Ref.~\citep{Marra2021} to alleviate the $H_0$ tension. In  \citep{Marra2021}, 
the $M$  variation is presumed to occur at $z\simeq 0.01$  and the value $\Delta M$, which corresponds to  parameter $A$,  is set to be  -0.2 in order to fully resolve the Hubble tension. 
In this paper, we  find a  sign for this transition from the SNe Ia data, and the absolute  value of $A$ is less than $0.2$.

Since a sudden transition of $M$ is a strong assumption, we now consider another ansatz that $M$ varies linearly with redshift  in the redshift region of $z_0\leq z<z_t$:
\begin{eqnarray}\label{eq:mod_M2}
	M(z)=\left\{\begin{array}{ll}
		M_0 & \mbox{ if } z< z_0 \\
		M_0+A\frac{z-z_0}{z_t-z_0} & \mbox{ if } z_0\leq z< z_t \\
		M_0+A & \mbox{ if } z\geq z_t \ ,
	\end{array} \right.
\end{eqnarray}
where $z_0$ and $z_t$ are two redshift points representing the beginning and ending of the variation of $M$, respectively.
We fix $z_0=0.01$ since the SNe Ia data used in our analysis satisfy  $z>0.01$. Thus, we have  the same free parameters ($A$ and $z_t$) as in the case of a sudden variation of $M$. From the Pantheon+ SNe Ia data, we obtain that the best fitting values are $\Omega_\mathrm{m0}=0.331$, $H_0= 68.19~\mathrm{ km~s^{-1}~Mpc^{-1}}$, $A=-0.156$, and $z_t= 0.0139$, with $\chi^2_\mathrm{min}=1395.6$.
Their mean values with 1$\sigma$ uncertainty are $\Omega_\mathrm{m0}=0.330\pm0.018$, $H_0=69.4^{+2.5}_{-2.0}~\mathrm{ km~s^{-1}~Mpc^{-1}}$, $A=-0.121^{+0.078}_{-0.052}$, and $z_t=0.0161^{+0.0016}_{-0.0047}$, respectively. 
Here,  the value of $\chi^2_\mathrm{min}$ is larger than that  obtained in  the case of $M$ changing suddenly, although it is still  smaller than the value in the constant $M$ case  for about $6.6$. The constraint on $\Omega_\mathrm{m0}$ is almost the same as  that in both  the cases of a constant and a suddenly varying  $M$. The mean value of $z_t$ becomes larger slightly than the one from a sudden variation of $M$. The mean  value of $A$ is smaller than $A=-0.084$ obtained in the case of $M$ varying suddenly, which leads to the value of $H_0$ being smaller  than the one obtained by using Eq.~(\ref{eq:mod_M1}) and consistent with the result from  the CMB observations within $1\sigma$. Thus, the $H_0$ tension is further alleviated when $M$ varies linearly with redshift.
These results can be seen clearly in Fig.~\ref{Fig:1} and Tab.~\ref{tab:2}.

To further compare the standard model ($M=$constant) with the models with a varying $M$, we consider the Akaike information criterion (AIC)\cite{Akaike1974,Akaike1981} and the Bayesian information criterion (BIC) \cite{Schwarz1978}, which are defined as $\mathrm{AIC}=2p-2\ln \mathcal{L}$ and $\mathrm{BIC}=p\ln N-2\ln \mathcal{L}$, respectively. Here $p$ is the number of free parameters, $N$ is the number of data points, and $\mathcal{L}\propto\exp\left(-\chi^2/2\right)$ is the likelihood function. 
The difference in the AIC(BIC) of a given model relative to the reference model can be calculated by using $\Delta \mathrm{AIC}(\mathrm{BIC})=\mathrm{AIC(BIC)}-\mathrm{AIC_{ref}(BIC_{ref})}$.  Here the model with a constant $M$ will be set as the reference model.  If $0<|\Delta\mathrm{AIC}|<2$, it is difficult to single out a better model, while $4 < |\Delta\mathrm{AIC}| <7$ means mild evidence against the model with the larger AIC, and $|\Delta\mathrm{AIC}|>10$ suggests strong evidence against the model with the larger AIC~\cite{Burnham2004}.  For the $\Delta \mathrm{BIC}$, a range of $0<|\Delta\mathrm{BIC}|<2$ also indicates difficulty in preferring the model, and  $2 < |\Delta\mathrm{BIC}| <6$ and $|\Delta\mathrm{BIC}|>6$ are regarded  positive and strong evidences, respectively,   against the model with the larger BIC~\cite{Jeffreys1998}.  We find that $\Delta\mathrm{AIC(BIC)}=-4.8(6)$ for the sudden transition $M$ model, and $\Delta\mathrm{AIC(BIC)}=-2.5(8.2)$ for the linear transition $M$ model.  It is apparent that the AIC mildly prefers the sudden $M$ transition model since its value is more than 4 less than that of the constant $M$ model, while the BIC still favors the standard model.

\begin{table}
	\caption{\label{tab:2}
		\footnotesize
		Marginalized Constraints on Parameters in Different Transition Models of $M$.
	}
	\centering 
	\scriptsize
	\begin{threeparttable}
		\begin{tabular}{c|ccc}
			\hline
			\hline
			\multicolumn{4}{c}{Pantheon+ Sample}\\ 
			\hline
			& ~Sudden Transition~ & ~Linear Transition~ & ~Constant~ \\
			\hline
			$\Omega_{\rm m0}$ & $0.332\pm0.018$ & $0.330\pm0.018$ & $0.333\pm0.018$\\
			$H_0$ & $70.5^{+2.0}_{-1.7}$ & $69.4^{+2.5}_{-2.0}$ & $73.22\pm0.95$  \\
			$A$ & $-0.084^{+0.061}_{-0.038}$ & $-0.121^{+0.078}_{-0.052}$ & -\\
			$z_t$ & $0.0139^{+0.0003}_{-0.0035}$ & $0.0161^{+0.0016}_{-0.0047}$ & -\\
			\hline
			$\Delta$AIC(BIC)$^{\rm b}$ & -4.8(6) & -2.5(8.2) & 0\\
			\hline
			\hline
			\multicolumn{4}{c}{ $f\sigma_8$ Sample }\\
			\hline
			$\Omega_{\rm m0}$ & $0.286^{+0.028}_{-0.033}$ & $0.285\pm0.031$ & $0.284^{+0.029}_{-0.032}$\\
			$\sigma_8$ & $0.801\pm0.021$ & $0.808^{+0.020}_{-0.023}$ & $0.774\pm0.020$  \\
			\hline
			$\Delta$AIC(BIC) & -0.04(-0.04) & -0.04(-0.04) & 0\\
			\hline
		\end{tabular}
		\begin{tablenotes}
			\item[a] The mean values with 1$\sigma$ uncertainty are shown.
			\item[b] $\Delta \mathrm{AIC}(\mathrm{BIC})=\mathrm{AIC(BIC)}-\mathrm{AIC_{ref}(BIC_{ref})}$, and the reference model is the constant $M$ model. 
		\end{tablenotes}
	\end{threeparttable}
\end{table}

\section{growth tension} \label{sec:4}
The absolute magnitude of a star is the star's luminosity when it is at a distance of $10$~pc. Thus, the difference between the post-transition SNe Ia absolute magnitude $M$ and the pre-transition $M_0$ can be connected with the absolute luminosity $L$ via $M-M_0=-\frac{5}{2}\log_{10}\frac{L}{L_0}$. 
Since  the peak luminosity of SNe Ia is determined by the mass of nickel synthesized ($m_\mathrm{Ni}$)~\cite{Arnett1982}, we  have a simple relation $L\propto m_\mathrm{Ni}\propto m_c$ after assuming  $m_\mathrm{Ni}$ is directly proportional to the Chandrasekhar mass $m_c$~\cite{Gaztanaga2001}, which can be estimated according to $m_c\simeq \frac{3}{m_e}\left(\frac{\hbar c}{G_\mathrm{eff}}\right)^{3/2}$, where $m_e$ is the mass per electron. Then, for a fixed $m_e$, one has $L\propto   G_\mathrm{eff}^{-3/2}$. 
Therefore, a change of $M$ found in the Sec.~\ref{sec:2} can be explained as a variation of $G_\mathrm{eff}$. Using  $\Delta \mu_G\equiv\mu_G-1$, where $\mu_G$ is defined as $\mu_G\equiv \frac{G_\mathrm{eff}}{G_N}$ with $G_N$ being the locally measured Newton's constant,  to denote the change of $G_\mathrm{eff}$ and   $\Delta M=M-M_0$, we obtain that 
\begin{eqnarray}
	\Delta \mu_G = 10^{\frac{4}{15}\Delta M}-1 .
\end{eqnarray}
Obviously, for the case of a sudden transition of $M$,  $\mu_G=1$ when $z<z_t$ and $\mu_G=1+\Delta \mu_G$ when $z\geq z_t$.  
Using the best fitting values of $A$, we have  $\Delta\mu_{G} -0.076$   and $-0.091$, respectively,  for the sudden and linear transition ansatzes of $M$ when $z\geq z_t$.
Apparently,  both values of $\Delta \mu_G$ are larger than $-0.12$, which is  derived from $\Delta M=-0.2$ assumed in~\cite{Marra2021}.

A variation of $\mu_G$ has an impact on the growth rate of the cosmological matter fluctuations $\delta(a)=\frac{\delta \rho}{\rho}(a)$ since the linear growth satisfies the equation: 
\begin{eqnarray}\label{eq:delta}
	\delta''+\left( \frac{3}{a} + \frac{H'(a)}{H(a)} \right)\delta'-\frac{3}{2}\frac{\Omega_\mathrm{m0}\mu_G}{a^5H(a)^2/H_0^2}\delta =0,
\end{eqnarray}
where $a=(1+z)^{-1}$ is the scale factor, a prime denotes a derivative with respect to $a$, $\rho$ is the matter density and $\delta\rho$ represents the fluctuation of matter density.  If one uses  parameter $\sigma_8$ to quantify the linear growth of the perturbations, the values of $\sigma_8$ and $\Omega_\mathrm{m0}$ derived from the measurements of  the weak lensing and galaxy redshift space distortions disagree at about $2-3\sigma$ level with  those inferred from the Planck  CMB data~\cite{Perivolaropoulos2022,Valentino2021,Abdalla2022}.  Now, we discuss what happens to  $\sigma_8$ and $\Omega_\mathrm{m0}$ when a modification of $\mu_G$ is introduced. 
To estimate their values, let us note that  a commonly observed measurement is the quantity $f\sigma_8$:
\begin{eqnarray}
f\sigma_8=\frac{\sigma_8}{\delta(a=1)}a\delta'(a,\Omega_\mathrm{m0},\mu_G).
\end{eqnarray}
To obtain $\delta$ and $\delta'$, we choose  the initial conditions to be $\delta(a_\mathrm{ini}\ll1)=a_{\rm ini}$ and $\delta'(a_\mathrm{ini}\ll1)=1$ with $a_\mathrm{ini}\sim 10^{-3}$~\cite{Nesseris2017}, and then numerically solve the linear growth equation (Eq.~(\ref{eq:delta})) by using the function \textit{scipy.integrate.odeint} in Python.
In our analysis, 62 observational $f\sigma_8$ data within a redshift range of $0.02\leq z \leq 1.944$ are used, which are collected in Ref.~\cite{Kazantzidis2018}.
The minimum $\chi^2$ method is also used here, which is expressed as:
\begin{eqnarray}
	\chi^2_{f\sigma_8}=\left [\bm{\hat{f\sigma_8}}_\mathrm{obs}-\frac{f{\sigma_8}_\mathrm{th}}{q}\right]^\dagger C^{-1}_{f\sigma_8} \left[\bm{\hat{f\sigma_8}}_\mathrm{obs}-\frac{f{\sigma_8}_\mathrm{th}}{q} \right].
\end{eqnarray}
Here $q$ is the correction factor, dependent on the referenced model of observational data~\cite{Kazantzidis2018}, and $C_{f\sigma_8}$ is the covariance matrix of $f\sigma_8$ sample. Since $M$ varies in $z<0.02$ and data are located at $z>0.02$, we thus fix the value of $\mu_G$ to be $1+\Delta \mu_G$  when the two ansatzes  of varying $M$ are considered.

\begin{figure}
	\centering
	\includegraphics[width=0.6\textwidth]{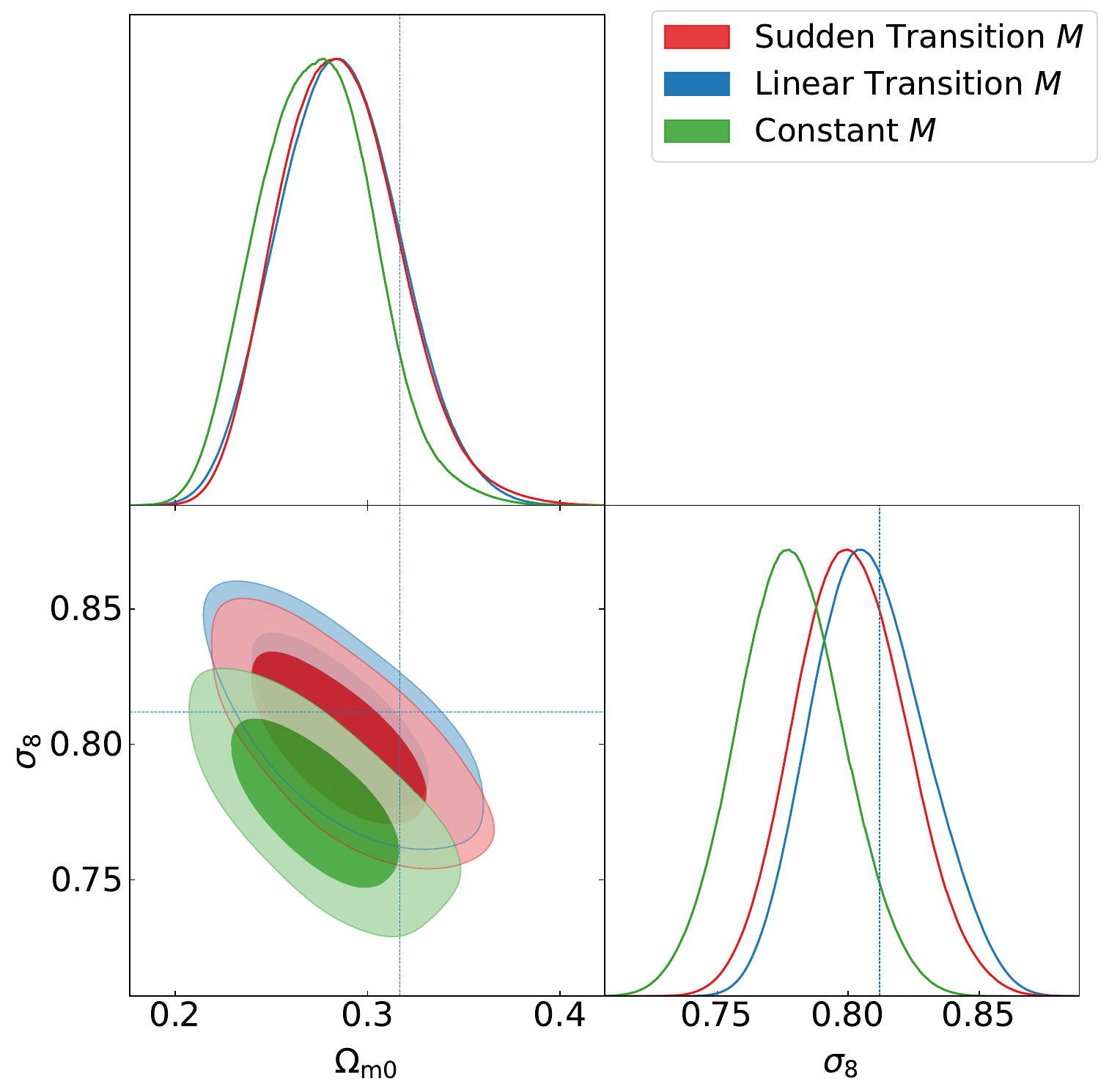}
	\caption{Constraints on $\Omega_\mathrm{m0}$ and $\sigma_8$ from the $f\sigma_8$ data. The blue dotted lines indicate the results from the Planck CMB data.
		\label{Fig:2}}
\end{figure}

We find that, when $\mu_G=1$, the mean values with 1$\sigma$ uncertainty of the parameters are $\Omega_\mathrm{m0}=0.284^{+0.029}_{-0.032}$ and  $\sigma_8=0.774\pm 0.020$ with $\chi^2_\mathrm{min}=29.016$.
The corresponding  contour plots on $\sigma_8-\Omega_\mathrm{m0}$ plane are shown in Fig.~\ref{Fig:2} and Tab.~\ref{tab:2}.  These vaules deviate significantly  from those given by the CMB data at more than 2$\sigma$ CL, and the marginalized constraint on $\sigma_8$ also deviates from the CMB's result.
When a modified $\mu_G$ is considered, we obtain  $\Omega_\mathrm{m0}=0.286^{+0.028}_{-0.033}$ and $\sigma_8=0.801\pm 0.021$ with $\chi^2_\mathrm{min}=28.976$ for the sudden transition ansatz, and $\Omega_\mathrm{m0}=0.285\pm 0.031$ and $\sigma_8=0.808^{+0.020}_{-0.023}$ with $\chi^2_\mathrm{min}=28.970$ for the linear transition one.
The contour plots of the constraints on $\Omega_\mathrm{m0}$ and $\sigma_8$ with both ansatzes, presented in Fig.~\ref{Fig:2}, are close to the CMB's result within about 1$\sigma$ CL, and the marginalized constraints on $\sigma_8$ are well-consistent with that from the  CMB data.
Thus, a modified $G_\mathrm{eff}$ does help  resolve the growth tension.
Setting the model with $\mu_G=1$ as the reference one, we also calculate the values $\Delta\mathrm{AIC(BIC)}$ for both ansatzes, and find that $\Delta\mathrm{AIC}=\Delta \mathrm{BIC}\simeq -0.04$ since all three models have the same free parameters. Therefore, a model favored by the $f\sigma_8$ observational data cannot be singled out by using the AIC and BIC.

\section{Conclusions} \label{sec:5}

We propose a cosmological-model-independent  method to obtain the apparent magnitude $m$ and its derivative $m'$ at different redshift points from the SNe Ia data, and find that the Pantheon+ sample supports  deviation of $m$ and $m'$ from the predictions of the $\Lambda$CDM model at the lowest redshift point.  This deviation may be explained as  a result of a  transition of the absolute magnitude $M$ in the low redshift region. The observations seem to support this transition since the minimum value of $\chi^2$ for two  ansatzes of a varying $M$ is less than that for a constant $M$. Furthermore, the AIC prefers the model with a sudden transition of $M$ although the BIC still supports the constant $M$ model.   With a varying $M$, the $H_0$ tension is alleviated to be about $1$-$2$ $\sigma$ and the growth tension can be resolved after attributing the variation of $M$ to a modification of the effective Newton's constant. 

The variation of $M$ or $G_\mathrm{eff}$ may indicate that the theory of general relativity needs to be extended~\cite{Wright2018,Lee2020,Sapone2021,Perivolaropoulos2022b,Escorcio2023,Camarena2023,Ruchika2023,Mukherjee2024,Gomez-Valent2024, Schiavone2023, Silva2024, Vagnozzi2023,Hogas2023}.  
If the strength of gravity or $M$ evolves over time at very low redshifts, the SNe Ia are no longer  standardizable candles, and  thus the cosmology implied by the existing SN Ia data will be different~\cite{Wright2018}.   
Moreover, a varying $G_\mathrm{eff}$ not only induces the change in $M$ but also affects the low redshift galaxy survey data~\cite{Alestas2022b}, the period-luminosity relation in the Cepheid~\cite{Sakstein2019,Ruchika2023}, as well as the expected fluxes of neutrinos and x-rays from neutron stars~\cite{Goldman2024}.

\begin{acknowledgements}
This work was supported in part by the NSFC under Grant Nos. 12275080 and 12075084, and the innovative research group of Hunan Province under Grant No. 2024JJ1006.
\end{acknowledgements}

\appendix
\section{}\label{sec:app}

To check the reliability of our method proposed in Sec. II, we plan to simulate the SNe Ia data to constrain   $m_i$ and $m'_i$.  The mock data are  generated through following processes:
Firstly, we apply the Kernel Density Estimate (KDE) with a band width $b=0.01$ to describe the redshift distribution of the Pantheon+ sample, and then use this redshift distribution to sample  randomly  the same redshift  points (1560 points) as the Pantheon+ sample in the redshift region of $0.01<z\leq 0.8$.  Secondly, at the every  redshift point, the value of $\langle m_\mathrm{th}\rangle$ is calculated from Eq.~(\ref{eq:m}) by assuming a fiducial model:  $\Lambda$CDM model with $\Omega_\mathrm{m0}=0.333$, $H_0=73.22~ \mathrm{km~s^{-1}Mpc^{-1}}$, and $M=-19.253~\mathrm{mag}$.
Thirdly, the mock $m_\mathrm{sim}$ can be sampled from $\mathcal{N}(\langle m_\mathrm{th}\rangle,\sigma_\mathrm{SN})$. Here $\mathcal{N}$ means the Gaussian distribution and  $\sigma_\mathrm{SN}$ is the uncertainty of  $\langle m_\mathrm{th}\rangle$, which is derived from the Pantheon+ sample by using   the KDE method. Thus, the mock 1560 SNe Ia data are obtained.

The constraints on $m_i$ and $m'_i$ from the mock data  can be achieved by using the minimum $\chi^2$ method (Eq.~(\ref{eq:chi})),  and  $\Delta m_i\equiv m_i-m_{i,\mathrm{th}}$ and $\Delta m'_i\equiv m'_i-m'_{i,\mathrm{th}}$ in each bin can be calculated, where the subscript `th' denotes the prediction from the fiducial model.
After repeating above process 1000 times, we can plot the distributions of the deviations $\Delta m_i$ and $\Delta m'_i$.
If these 1000 deviations in each bin are concentrated around $\Delta m_i=0$ and $\Delta m'_i=0$, it implies that our method is feasible.

\begin{figure}
	\includegraphics[width=0.45\textwidth]{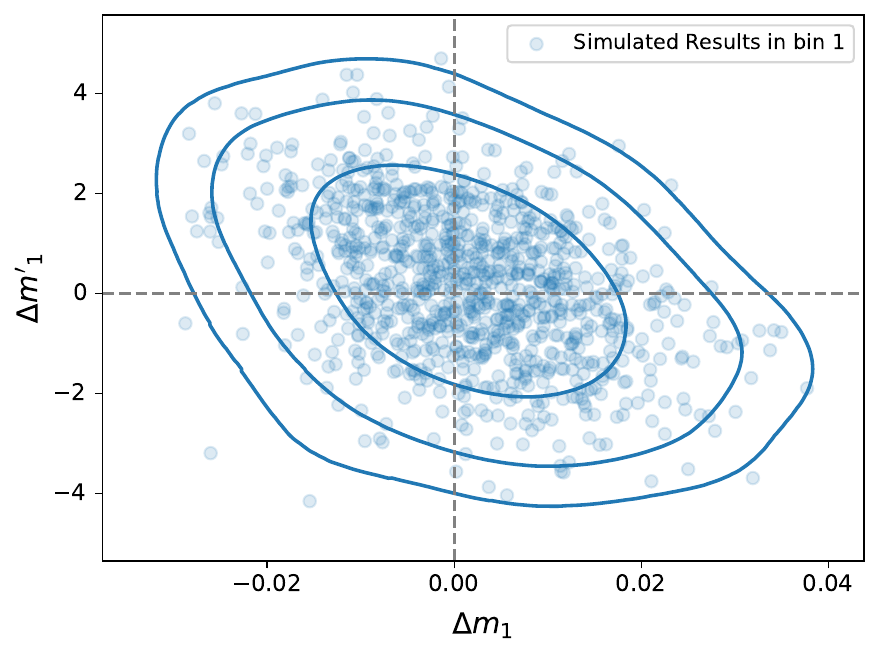}
	\includegraphics[width=0.45\textwidth]{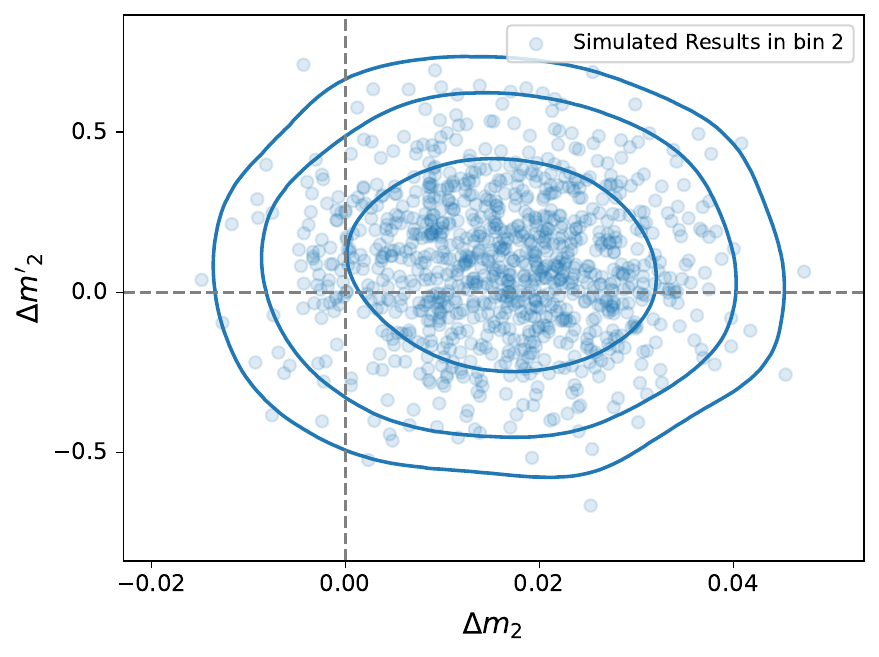}
	\includegraphics[width=0.45\textwidth]{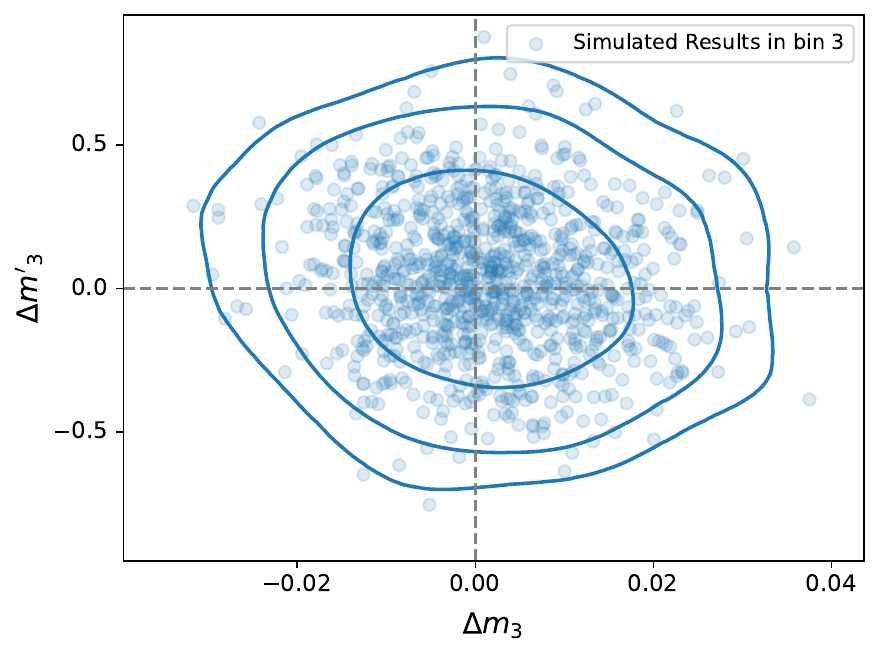}
	\includegraphics[width=0.45\textwidth]{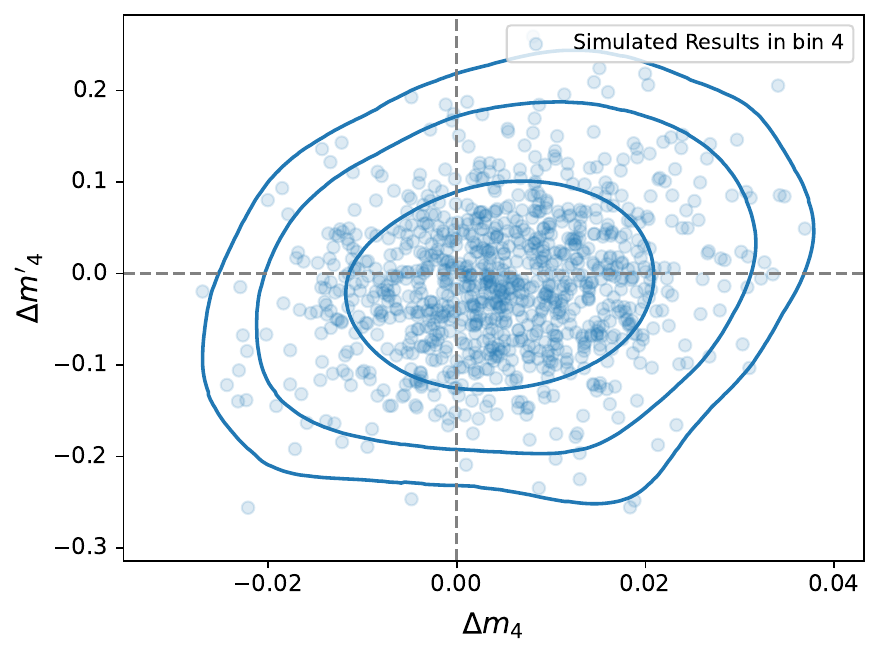}
	\caption{
		The distribution of $\Delta m_i$ and $\Delta m'_i$ from 1000 time simulated data (blue points) in the case of four bins.
		The blue solid lines are the 68\%, 95\%, and 99\% confidence levels, respectively.
		\label{Fig:3}
	}
\end{figure}

\begin{figure}
	\includegraphics[width=0.45\textwidth]{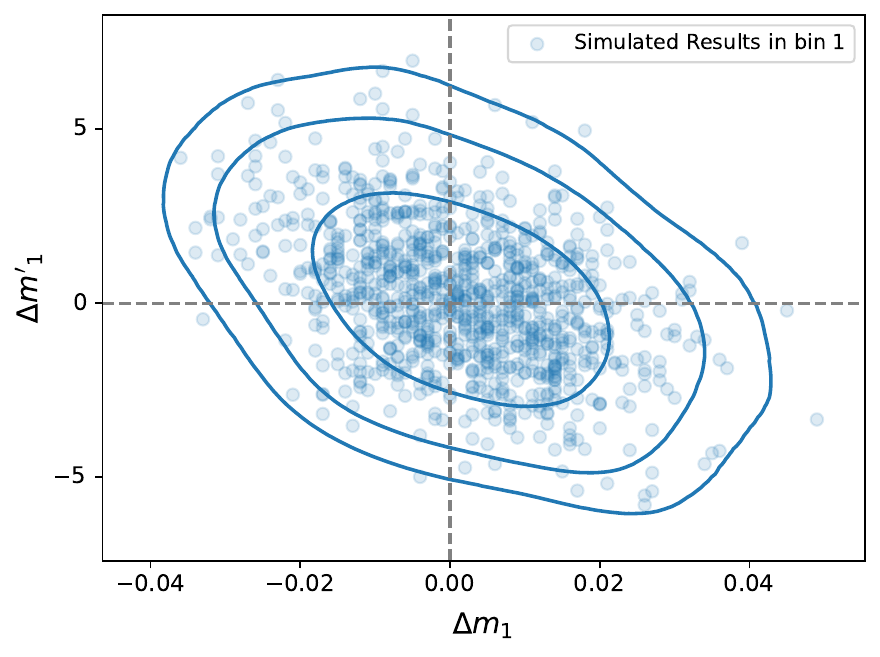}
	\includegraphics[width=0.45\textwidth]{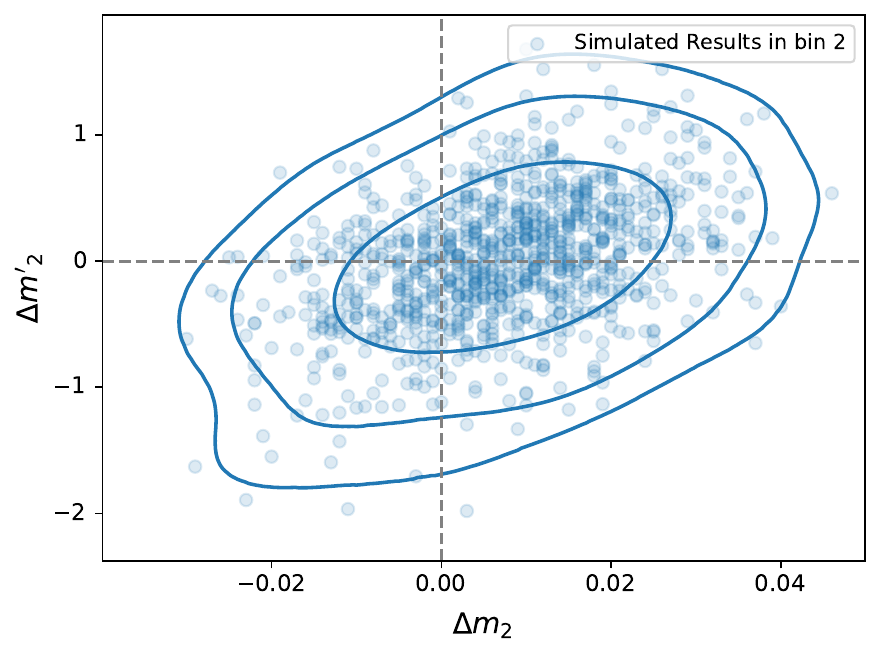}
	\includegraphics[width=0.45\textwidth]{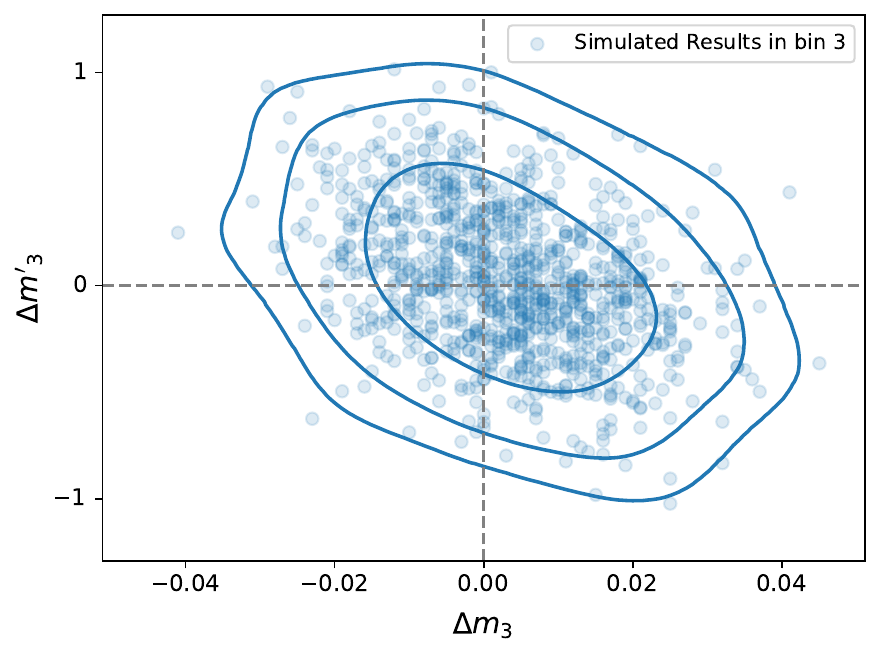}
	\includegraphics[width=0.45\textwidth]{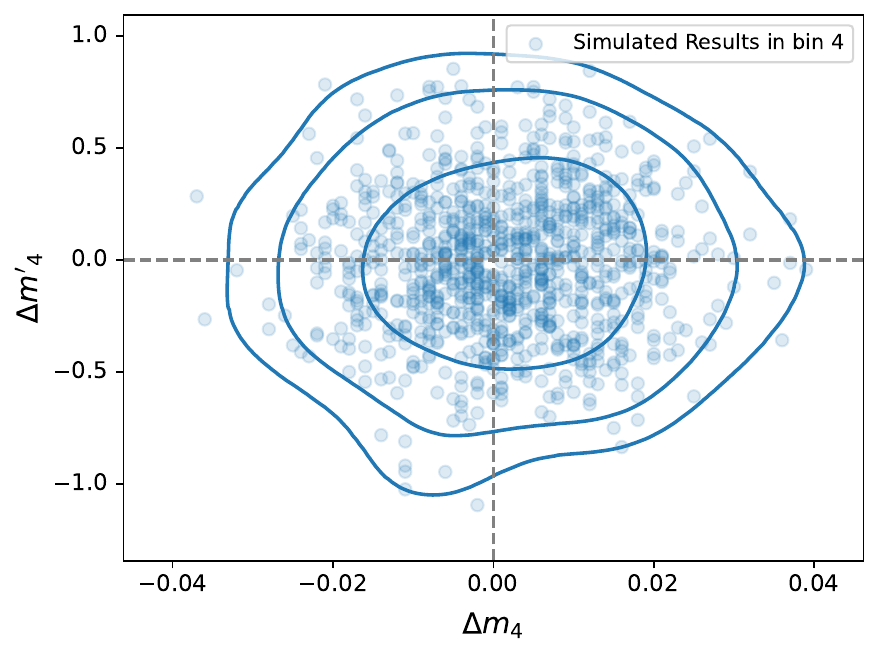}
	\includegraphics[width=0.45\textwidth]{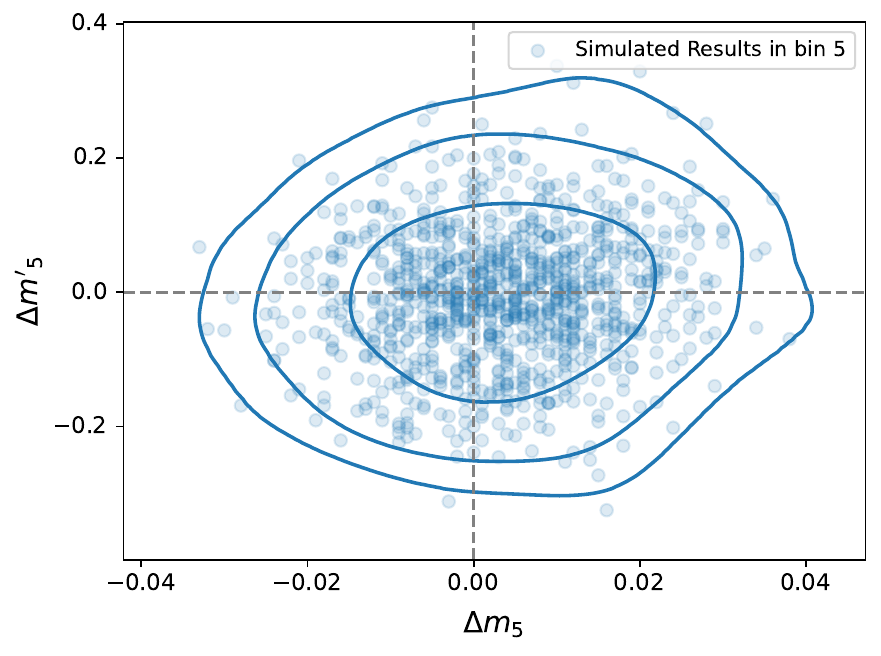}
	\caption{
		The distribution of $\Delta m_i$ and $\Delta m'_i$ from 1000 time simulated data (blue points) in the case of  five bins.
		The blue solid lines represent the 68\%, 95\%, and 99\% confidence levels, respectively.
		\label{Fig:4}
	}
\end{figure}

\begin{figure}
	\includegraphics[width=0.45\textwidth]{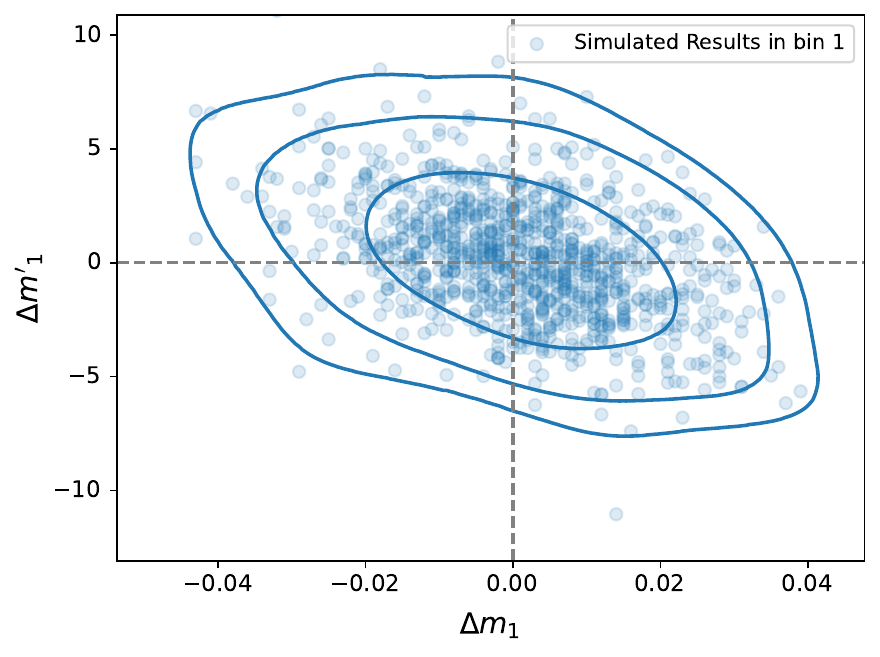}
	\includegraphics[width=0.45\textwidth]{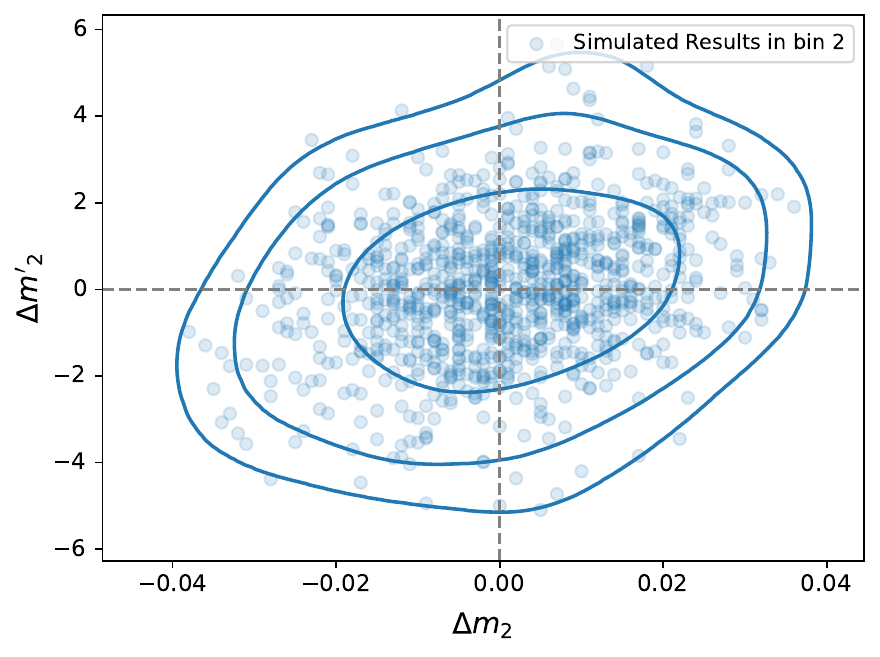}
	\includegraphics[width=0.45\textwidth]{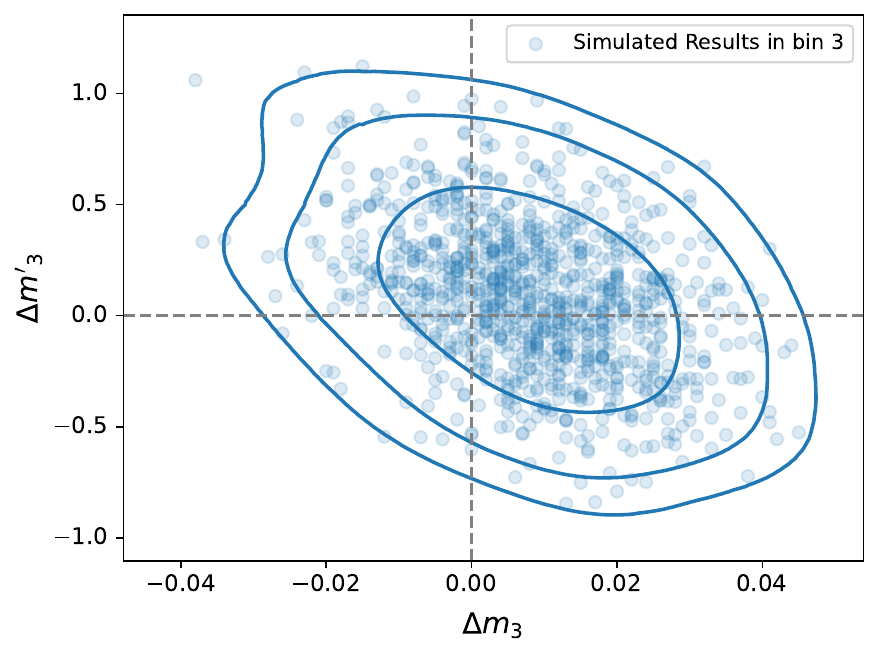}
	\includegraphics[width=0.45\textwidth]{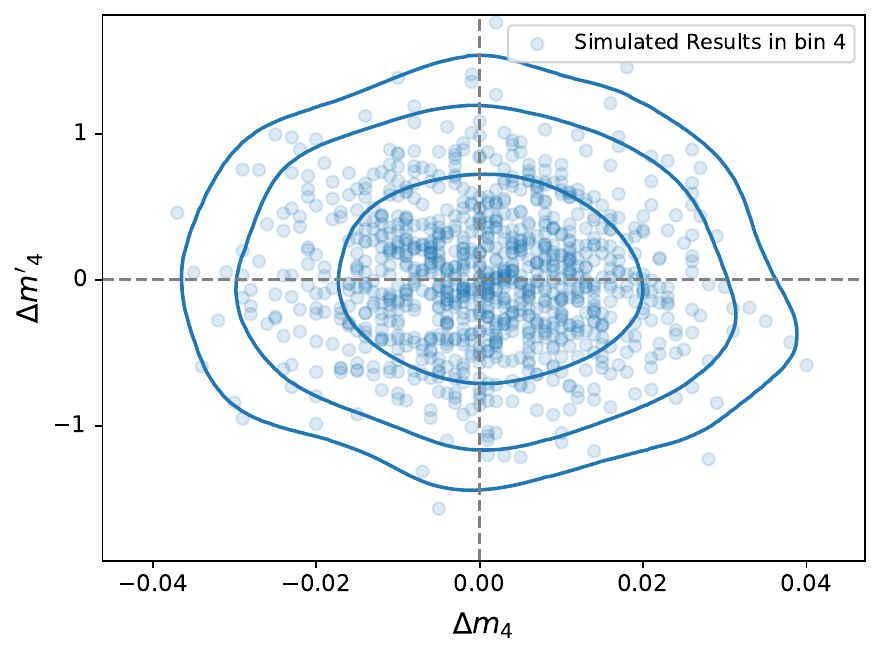}
	\includegraphics[width=0.45\textwidth]{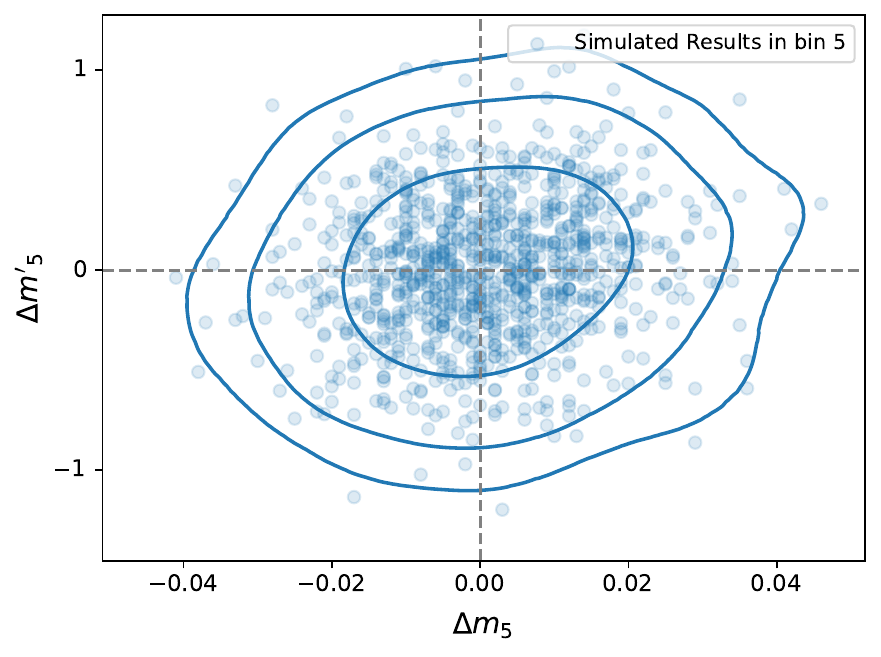}
	\includegraphics[width=0.45\textwidth]{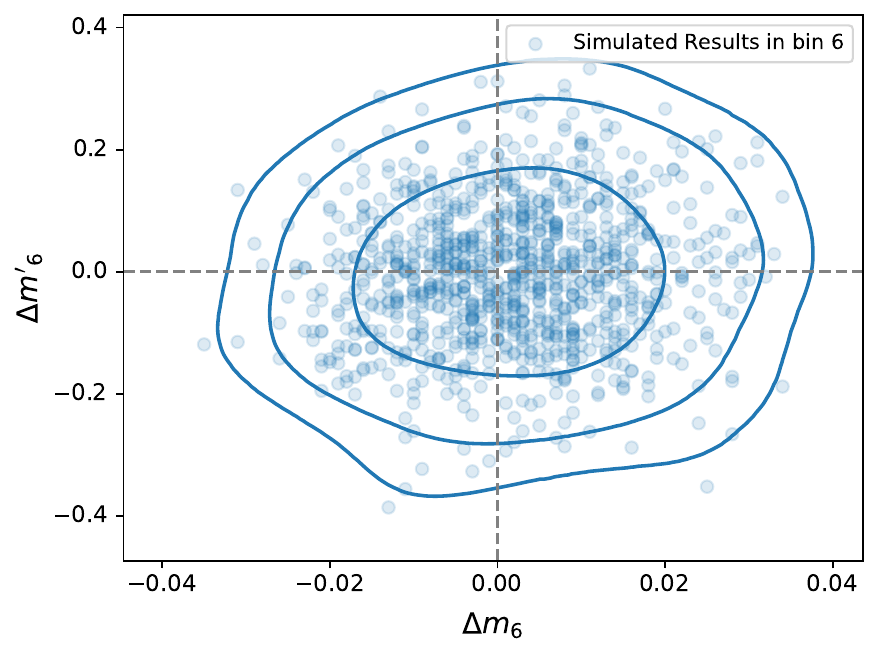}
	\caption{
The distribution of $\Delta m_i$ and $\Delta m'_i$ from 1000 time simulated data (blue points) in the case of  six bins.
		The blue solid lines represent the 68\%, 95\%, and 99\% confidence levels, respectively.		\label{Fig:5}
	}
\end{figure}

We consider three different cases: dividing the mock data into four, five, and six bins with the same number data of points in each bin, respectively.
Figures~\ref{Fig:3}, \ref{Fig:4}, and \ref{Fig:5} show the results of four, five, and six bins, respectively.
From Fig.~\ref{Fig:3}, one can see that the mean of 1000 $\Delta m_2$ deviates from zero at about $1\sigma$ confidence level, although   the results in other bins are compatible  with the fiducial model.  Figures~\ref{Fig:4}, and \ref{Fig:5} indicate that when the number of bin is larger than 4,   all results  are consistent with the fiducial model very well. Thus, we  find  that the results from the Pantheon+ sample will be reliable since five bins are chosen in our analysis.

\section{}\label{sec:app2}

Here we study the constraints on $m_i$ and $m'_i$ from the Pantheon+ data  in the cases of four and six bins. The results are shown in Table~\ref{tab:3} and Table~\ref{tab:4}, respectively. In the case of four bins, we find that $m_i$ and $m'_i$ with $i>2$ are consistent with the prediction of the fiducial model, $\Delta m_2$ and $\Delta m'_2$ are compatible with zero only in $2\sigma$ CL, which are similar to the results from the mock data (see Fig.~\ref{Fig:3}),  while $m_1'$ deviates from that in the $\Lambda$CDM model at more than $2\sigma$ CL.   When six bin is considered, we find that only $\Delta m'_1$ is inconsistent with zero at more than $2\sigma$ CL, which is similar to what are obtained in the four and five bin cases.

\begin{table}
	\caption{\label{tab:3}
		\footnotesize
		Expanding Redshift Point $z_i$, Number of SNe~Ia, and Constraints on $m_i$ and $m'_i$ in the four bin case.
	}
	\centering
	\scriptsize
	\begin{threeparttable}
		\begin{tabular}{c|cccc}
			\hline
			\hline
			& bin 1 & bin 2 & bin 3 & bin 4 \\
			\hline
			$z_i$ & 0.018 & 0.075 & 0.244 & 0.513 \\
			~Redshift range~ & ~$0.010<z\leq0.031$~ & ~$0.031<z\leq0.181$~ & ~$0.181<z\leq0.330$~ & ~$0.330<z\leq0.799$~ \\
			$N_\mathrm{SN}$ & 390 & 390 & 390 & 390 \\
			\hline
			$m_i$ & $15.097\pm0.013$ & $18.327\pm0.007$ & $21.089\pm0.008$ & $22.943\pm0.010$  \\
			$m'_i$ & $119.979\pm1.883$ & $29.978\pm0.166$ & $10.132\pm0.155$ & $4.953\pm0.063$  \\
			\hline
			$\Delta m_i$  & $0.015\pm0.015$ & $0.015\pm0.010$ & $0.009\pm0.013$ & $-0.011\pm0.018$ \\
			$\Delta m'_i$  & $-4.023\pm1.883$ & $-0.267\pm0.168$ & $0.088\pm0.158$ & $-0.094\pm0.066$ \\
			\hline
		\end{tabular}
		\begin{tablenotes}
			\item[a] The mean values with 1$\sigma$ uncertainty are shown.
			\item[b] $\Delta m_i$($\Delta m'_i$) denote the differences between the constraints on $m_i$($m'_i$) and the fiducial model: $\Lambda$CDM model with $\Omega_\mathrm{m0}=0.333\pm 0.018$ and $\mathcal{M}=25+5\log_{10}\left(\frac{c}{H_0}\right)+M=23.808\pm0.007$. 
		\end{tablenotes}
	\end{threeparttable}
\end{table}
\begin{table}
	\caption{\label{tab:4}
		\footnotesize
		Expanding Redshift Point $z_i$, Number of SNe~Ia, and Constraints on $m_i$ and $m'_i$ in the six bin case.
	}
	\centering
	\scriptsize
	\begin{threeparttable}
		\begin{tabular}{c|cccccc}
			\hline
			\hline
			& bin 1 & bin 2 & bin 3 & bin 4 & bin 5 & bin 6 \\
			\hline
			$z_i$ & 0.0157 & 0.034 & 0.092 & 0.223 & 0.339 & 0.577 \\
			Redshift range & $0.010<z\leq0.025$ & $0.025<z\leq0.047$ & $0.047<z\leq0.181$ & $0.181<z\leq0.275$ & $0.275<z\leq0.417$ & $0.417<z\leq0.799$   \\
			$N_\mathrm{SN}$ & 260 & 260 & 260 & 260 & 260 & 260 \\
			\hline
			$m_i$ & $14.840\pm0.016$ & $16.521\pm0.010$ & $18.767\pm0.008$ & $20.862\pm0.009$ & $21.902\pm0.009$ & $23.237\pm0.013$  \\
			$m'_i$ & $132.991\pm3.288$ & $67.606\pm1.557$ & $24.947\pm0.198$ & $10.545\pm0.306$ & $7.367\pm0.243$ & $4.461\pm0.091$ \\
			\hline
			$\Delta m_i$  & $0.031\pm0.018$ & $0.007\pm0.012$ & $-0.0004\pm0.0112$ & $0.002\pm0.013$ & $0.009\pm0.015$ & $-0.021\pm0.021$ \\
			$\Delta m'_i$  & $-7.134\pm3.289$ & $2.034\pm1.557$ & $-0.078\pm0.200$ & $-0.372\pm0.308$ & $-0.059\pm0.244$ & $-0.059\pm0.093$ \\
			\hline
		\end{tabular}
		\begin{tablenotes}
			\item[a] The mean values with 1$\sigma$ uncertainty are shown.
			\item[b] $\Delta m_i$($\Delta m'_i$) denote the differences between the constraints on $m_i$($m'_i$) and the fiducial model: $\Lambda$CDM model with $\Omega_\mathrm{m0}=0.333\pm 0.018$ and $\mathcal{M}=25+5\log_{10}\left(\frac{c}{H_0}\right)+M=23.808\pm0.007$. 
		\end{tablenotes}
	\end{threeparttable}
\end{table}

\end{document}